\documentclass[journal=jpcc,manuscript=article]{achemso}
\usepackage{chemformula} 
\usepackage[T1]{fontenc} 
\usepackage{xcolor}
\usepackage{hyperref}

\author{Sushant Kumar Behera}
\affiliation{Materials Engineering, Indian Institute of Science Bengaluru 560012, India.}
\alsoaffiliation{Materials Science Division, Lawrence Berkeley National Laboratory Berkeley, CA 94720, USA.}

\author{Aparna Swain}
\affiliation{Department of Physics, Indian Institute of Science Bengaluru 560012, India.}
\alsoaffiliation{Department of Chemistry, University of Pennsylvania
Philadelphia, PA 19104, USA.}

\author{Praveen C Ramamurthy}
\email{praveen@iisc.ac.in}
\affiliation{Materials Engineering, Indian Institute of Science Bengaluru 560012, India}

\title[An \textsf{achemso} demo]
  { Lattice-Driven Electronic Structure Reconstruction in the Commensurate CDW Phase of 1T-Ta$S_2$ }


\begin{document}


\begin{abstract}
We investigate the structural and electronic reconstruction associated with the commensurate charge-density-wave (CCDW) phase in bulk and monolayer 1T-TaS$_2$ using density functional theory (DFT) and Wannier-based tight-binding modeling. Structural relaxation of a $\sqrt{13}\times\sqrt{13}$ supercell leads spontaneously to the formation of the Star-of-David (SoD) distortion, consistent with phonon softening of the undistorted phase. We focus on establishing a direct connection between real-space lattice distortion and momentum-space electronic reconstruction. Using Wannier interpolation, we demonstrate how the CCDW-induced Brillouin zone reduction leads to band folding, narrowing of Ta $5d$ bands, and reconstruction of the Fermi surface. Our analysis shows that features often interpreted as Fermi surface nesting emerge naturally from band folding associated with lattice distortion. We compare our calculated electronic structure with previously reported angle-resolved photoemission spectroscopy (ARPES) results at a qualitative level. While we do not explicitly compute electronic susceptibility or electron–phonon coupling matrix elements, the results provide a consistent microscopic framework linking lattice instability and electronic structure reconstruction in 1T-TaS$_2$.

\end{abstract}

\maketitle

\textit{Introduction}-~Transition metal dichalcogenides (TMDs), a class of two-dimensional (2D) layered materials, exhibit a wide range of electronic ground states spanning insulating, semiconducting, metallic, and superconducting phases \cite{PhysRevLett.117.106801, PhysRevLett.125.186401, PhysRevLett.125.165302}. Among these, charge density waves (CDWs) are particularly intriguing collective states that arise from an interplay between lattice instabilities and electronic degrees of freedom. Despite decades of study, the microscopic driving mechanism behind CDW formation in layered TMDs—especially commensurate CDW (CCDW) phases—remains an open question \cite{PhysRevLett.100.036804,PhysRevLett.129.156401}.

Early theoretical models attributed CDW formation primarily to favorable Fermi surface nesting (FSN), wherein parallel regions of the Fermi surface are connected by a characteristic wave vector, enhancing the electronic susceptibility and promoting a density modulation. However, extensive experimental and first-principles studies have demonstrated that FSN alone is often insufficient to explain the observed CCDW ordering vectors and transition temperatures in layered TMDs. Instead, these studies indicate that CDW formation is predominantly driven by lattice instabilities associated with strong electron–phonon coupling and periodic lattice distortions (PLDs), while FSN emerges as a secondary consequence of the reconstructed electronic structure. In this picture, the distortion-induced band folding and orbital hybridization reshape the Fermi surface, giving rise to apparent nesting features only after the lattice symmetry is lowered. The persistence of interaction-renormalized electronic states, enabled by reduced scattering and atomically sharp interfaces in low-dimensional materials, further stabilizes long-range CDW order \cite{PhysRevLett.130.226401,PhysRevLett.120.037002,PhysRevLett.122.086402}. Related 2D systems, including graphene-based heterostructures, provide additional evidence that reduced dimensionality enhances coupling effects such as superconductivity, magnetism, and spin–orbit interactions \cite{PhysRevB.77.033306,Behera_2021}.

A prototypical system exhibiting these competing mechanisms is 1T-Ta$S_2$, which hosts a remarkably rich CDW phase diagram involving multiple incommensurate, nearly commensurate, and fully commensurate phases. In particular, the low-temperature CCDW phase is characterized by a $\sqrt{13}\times\sqrt{13}$ periodic lattice distortion forming Star-of-David (SOD) clusters in the Ta layer. While photoemission and scattering experiments have revealed pronounced band reconstruction, partial gap opening, and Fermi surface reorganization across the CDW transition, the extent to which these features originate from FSN versus lattice-driven electronic reconstruction remains actively debated. First-principles studies of related layered CDW materials, such as 1T-Ta$S_2$, 1T-Ta$Se_2$, and 2H-NbSe$_2$, have shown that CDW ordering vectors often do not coincide with dominant nesting vectors, further supporting a lattice-driven mechanism. At the same time, the CCDW phase in 1T-Ta$S_2$ has been discussed as a correlation-enhanced localization state, where narrowed Ta~5$d$ bands amplify interaction effects without requiring FSN as the primary instability \cite{PhysRevLett.106.136809,Keimer2015,PhysRevB.107.L201109,C8CP04684K}.

Experimentally, techniques such as angle-resolved photoemission spectroscopy (ARPES), scanning tunneling microscopy/spectroscopy (STM/STS), and X-ray scattering have provided compelling evidence of CDW-induced electronic reconstruction in 1T-Ta$S_2$. ARPES measurements reveal band folding, spectral weight redistribution, and partial gap opening at the Fermi level in the CCDW phase, while STM/STS studies uncover localized electronic states associated with the SOD clusters \cite{Chen2015,pnas.1211387110}. These observations highlight the importance of considering not only the reconstructed band structure but also photoemission matrix-element effects when interpreting Fermi surface features and nesting signatures.

In this work, we investigate the structural and electronic origins of the CCDW phase in bulk and monolayer 1T-Ta$S_2$ using first-principles density functional theory combined with phonon calculations and Wannier-based tight-binding modeling. We demonstrate that soft phonon modes of the normal phase signal an intrinsic lattice instability, and that structural relaxation along these modes spontaneously generates the CCDW Star-of-David distortion. The resulting periodic lattice distortion drives a pronounced reconstruction of the Ta~5$d$ electronic states, with Fermi surface nesting appearing as an emergent consequence of band folding rather than the primary driving force. By explicitly incorporating Wannier-based matrix-element effects, our study provides a consistent microscopic framework for understanding CCDW formation and electronic reconstruction in low-dimensional TMDs.

\textit{Methodology Set-ups}-~The structural optimization and electronic structure calculations of 1T-Ta$S_2$ were performed within the framework of density functional theory (DFT) using the Quantum ESPRESSO package \cite{PhysRevB.47.558,PhysRevB.49.14251,PhysRevB.54.11169}. The electron–ion interactions were described using ultrasoft pseudopotentials within a plane-wave basis set. Exchange–correlation effects were treated using the Perdew–Burke–Ernzerhof generalized-gradient approximation revised for solids (PBEsol) \cite{PhysRevB.79.155107}, which provides an improved description of equilibrium lattice parameters in layered materials. Commensurate charge-density-wave (CCDW) structures were constructed using a $\sqrt{13}\times\sqrt{13}$ supercell of the 1T-Ta$S_2$ primitive cell, initialized from experimentally reported periodic lattice distortion (PLD) atomic positions \cite{PhysRevB.56.13757}. For monolayer calculations, a vacuum spacing of 20~\AA\ was introduced along the out-of-plane ($c$) direction to eliminate spurious interactions between periodic images. Long-range dispersion interactions were included using the nonlocal van der Waals density functional rev-vdW-DF2 \cite{dft-d3,dft-d3-1}, which is essential for accurately capturing interlayer coupling and low-symmetry distorted configurations. For CCDW calculations, lattice parameters were fixed to isolate internal periodic lattice distortions; the rev-vdW-DF2 functional primarily ensures consistency between bulk and monolayer reference structures.

All structures were fully relaxed using the Methfessel–Paxton smearing scheme with a smearing width of $\sigma = 0.02$~eV until the residual Hellmann–Feynman forces on each atom were less than 0.002~eV/\AA\ and the total energy convergence threshold reached $10^{-6}$~eV. In the CCDW phase, atomic positions were fully relaxed while lattice parameters were fixed to those of the pristine high-symmetry phase to isolate the effects of internal periodic lattice distortions. A plane-wave kinetic energy cutoff of 45~Ry (corresponding to approximately 580~eV) was employed for the wavefunctions, with a proportionally higher charge-density cutoff, ensuring convergence of total energies and forces. Brillouin-zone integrations were carried out using a $\Gamma$-centered $12\times12\times4$ Monkhorst–Pack $k$-point mesh for structural relaxations and self-consistent electronic calculations of the primitive cell, while a denser $24\times24\times8$ mesh was used for density-of-states (DOS) calculations. The tetrahedron method was employed for accurate evaluation of the DOS and total energies.

Phonon dispersion relations were calculated using the finite-displacement method as implemented in the PHONOPY package \cite{phonopy}, interfaced with Quantum ESPRESSO, to identify lattice instabilities and soft phonon modes associated with the CCDW transition. To analyze Fermi-surface nesting, electronic reconstruction, and photoemission matrix-element effects, maximally localized Wannier functions (MLWFs) were constructed from Ta $5d$ orbitals using the WANNIER90 package \cite{MOSTOFI2008685}. The resulting Wannier-based tight-binding Hamiltonian was employed to compute band dispersions, Fermi-surface contours, and simulated angle-resolved photoemission spectra, which were compared directly with experimentally reported matrix-element-resolved ARPES data \cite{PhysRevLett.122.127001}. We note that the present work primarily focuses on bulk and monolayer electronic structures. While surface spectral features can be sensitive to termination and reconstruction, a detailed analysis of surface states is beyond the scope of this study and is not discussed further.

\textit{Tight-binding model and Wannierization}-~The tight-binding Hamiltonian was constructed based on maximally localized Wannier functions (MLWFs) obtained from first-principles calculations. Wannierization was performed by projecting exclusively onto Ta~5$d$ orbitals, which dominate the low-energy electronic structure near the Fermi level in both the normal and CCDW phases of 1T-Ta$S_2$. The Wannier disentanglement window was chosen to span the Ta~5$d$-derived bands within ±2 eV around the Fermi level, ensuring faithful reproduction of the low-energy electronic structure. Sulfur $3p$ orbitals were not explicitly included in the Wannier basis; their contribution enters implicitly through Ta–S hybridization encoded in the effective Ta–Ta hopping parameters derived from the Wannierization procedure. In constructing the tight-binding Hamiltonian, we retained the full Slater–Koster form of Ta–Ta hopping integrals, including $dd\sigma$, $dd\pi$, and $dd\delta$ channels, following Ref.~\cite{PhysRevB.73.073106}. In an undistorted octahedral environment, these hopping amplitudes are commonly assumed to scale asymptotically as $R^{-5}$ with the Ta–Ta interatomic distance. However, in the commensurate CDW phase of 1T-Ta$S_2$, this scaling significantly underestimates the effective hopping strengths due to the presence of large in-plane Ta displacements, shortened Ta–Ta bond lengths, and enhanced Ta–S–Ta hybridization induced by the Star-of-David distortion.

To capture the effective range of Ta–Ta and Ta–S–Ta hybridization, we introduce a distance-dependent scaling of the Slater--Koster hopping integrals proportional to $R^{-3}$. This form reflects the rapid decay of orbital overlap in transition-metal systems and is used as an effective parametrization rather than a fundamental asymptotic law. The scaling parameters are determined by fitting to Wannier90-extracted hopping amplitudes, enabling accurate reproduction of the relative strengths of nearest- and next-nearest-neighbor interactions, as well as the reconstructed band dispersions and Fermi surface topology obtained from DFT. We emphasize that this parametrization is empirical and primarily intended to describe low-energy electronic structure. Its transferability to other properties, such as phonons or electronic susceptibilities, is not established within the present work. 

Consistent with previous experimental and theoretical reports, the in-plane displacements of Ta atoms in the CCDW phase are approximately twice as large as the corresponding out-of-plane displacements of S atoms \cite{Rossnagel2011}, further justifying the dominance of modified in-plane Ta–Ta hopping in the low-energy electronic reconstruction. For the high-temperature normal phase, the primitive unit cell contains one Ta atom and two S atoms, resulting in a $6\times6$ tight-binding Hamiltonian when only Ta~5$d$ orbitals are retained. In contrast, the $\sqrt{13}\times\sqrt{13}$ CCDW supercell comprises 13 Ta atoms and 26 S atoms, yielding a $78\times78$ Hamiltonian matrix. In the present work, SOC was consistently neglected in both DFT and Wannier-based tight-binding calculations, reducing the effective Hamiltonian size accordingly. Moreover, SOC primarily induces band splittings away from the Fermi level and does not qualitatively affect the lattice-driven CCDW instability or Fermi surface topology at the energy scales relevant here. The energy eigenvalues, band dispersions, constant-energy contours, and Fermi surface maps were obtained by direct diagonalization of the tight-binding Hamiltonian. Slater–Koster hopping parameters were initialized from literature values \cite{PhysRevB.73.073106} and subsequently rescaled according to the inverse cubic Ta–Ta distance to achieve quantitative agreement with the Wannier-interpolated bands and the simulated photoemission matrix-element effects.

\textit{Results and Discussion}-~Bulk 1T-TaS$_2$ enters the commensurate CDW (CCDW) phase below 170 K, making direct experimental determination of the CCDW lattice parameters challenging \cite{PhysRevB.105.L081106}. To model this phase, we construct a $\sqrt{13}\times\sqrt{13}$ supercell containing 13 Ta and 26 S atoms and perform full structural relaxation of both atomic positions and lattice parameters starting from a slightly perturbed high-symmetry configuration. Upon relaxation, the system spontaneously stabilizes the Star-of-David (SOD) distortion, confirming an intrinsic lattice instability of the undistorted phase. The optimized supercell lattice constant is 11.98 \AA, and the shortest Ta–Ta bond within the SoD cluster is 3.135 \AA, reduced from 3.317 \AA\ in the high-symmetry phase.
\begin{figure*}
\centering
\includegraphics[width=12.5cm,height=10.0cm]{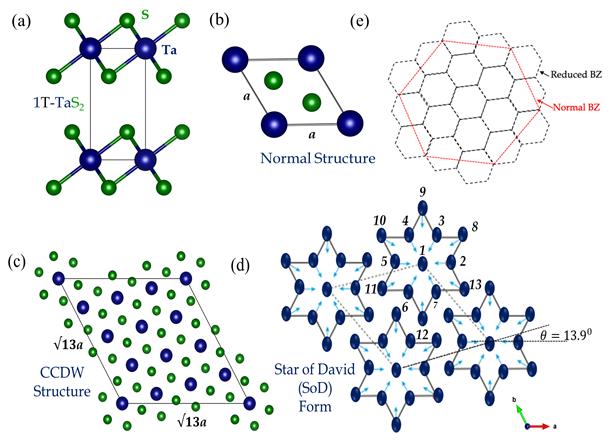}
  \caption{Optimized geometry of (a) normal phase, (b) unit cell of 1T-Ta$S_2$. (c) $\sqrt{13}\times\sqrt{13}$ superlattice unit cell of CCDW phase and (d) Star of David (SOD) formation with periodic random lattice distortions. (e) shows the size of Brillouin zones of normal (dotted red line) and CCDW reduced reciprocal cases (dotted black lines).}
  \label{fig-1}
\end{figure*}

\begin{figure*}
\centering
\includegraphics[width=12.0cm,height=12.5cm]{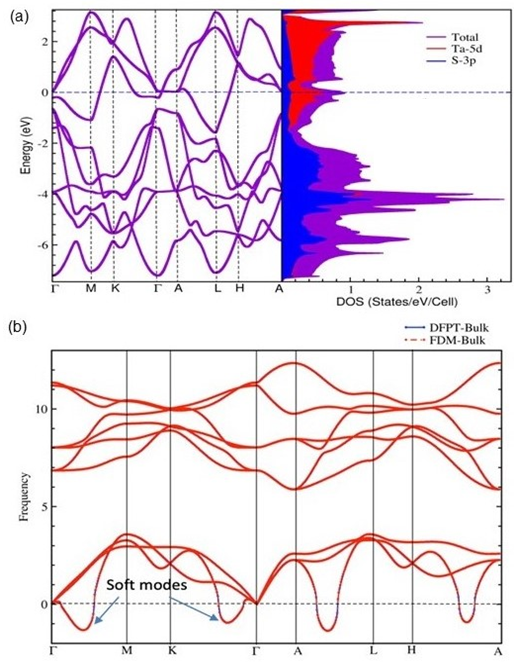}
  \caption{Electronic structures of a 1T-Ta$S_2$ normal bulk phase (a) the band structure along high symmetry directions. Side view, the total and partial DOS plot. (b) phonon dispersion plot showing soft phonon modes (\textit{negative frequency}) in blue arrow heads.}
  \label{fig-2}
\end{figure*}

\begin{figure*}
\centering
\includegraphics[width=9.0cm,height=8.0cm]{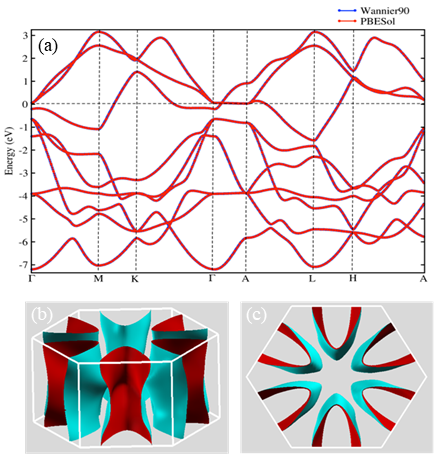}
  \caption{(a) Electronic band structure of monolayer 1T-Ta$S_2$ in the normal phase, where red curves denote DFT (PBESol) bands and blue curves denote Wannier90-interpolated bands, demonstrating excellent agreement. (b,c) Fermi surface contours obtained from the Wannier tight-binding Hamiltonian shown from side and top views, respectively.}
  \label{fig-3}
\end{figure*}

\begin{figure*}
\centering
\includegraphics[width=12.5cm,height=7.0cm]{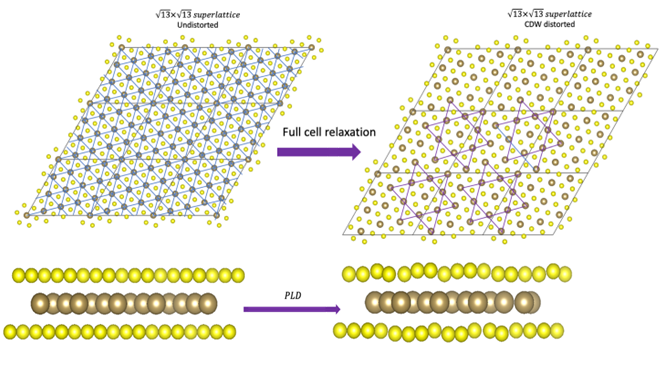}
  \caption{$\sqrt{13}\times\sqrt{13}$ superlattice unit cell of undistorted 1T-Ta$S_2$ monolayer and full cell volume relaxed distorted superlattice unit cell. The middle panel shows the atomic level \textit{periodic lattice distortions}.}
  \label{fig-4}
\end{figure*}

\textit{Commensurate Charge Density Wave States}: The CDW in 1T-Ta$S_2$ can exist in various commensurate phases, characterized by different periodicity. These phases include 4a, 4a+b, 5a, 6a, and more complex superlattice structures. Each commensurate CDW phase exhibits distinct electronic properties and ordering patterns. In this regard, x-ray scattering experiments revealed the formation of a 5a CDW phase with a periodicity of five times the lattice constant, leading to the formation of an unusual quasi-periodic superlattice \cite{PhysRevB.107.165115}. The CDW-induced structural reconstruction helps to observe the electronic modulations. Our obtained band structures for bulk and monolayer CCDW 1T-Ta$S_2$ phases are shown in Figure \ref{fig-2} and Figure \ref{fig-3}, respectively. The resulting minimizing forces and stress of structure relaxation are reconstructed as SOD phase. It shows the unit cell of the \textit{periodic lattice distortions (PLD)} and SOD clusters in the Ta plane (shown in Figure \ref{fig-4}). Here, the commensurate phase of the CDW phase can only be computationally accessible for periodic DFT simulations, thus we focus of the commensurate-CDW (CCDW) phase. Moreover, it is noticed that the superlattice unit cell contains maximum three types of non-equivalent Ta atoms. Here, a=$a_0$+3$b_0$ and b=4$a_0$-$b_0$ and c=$c_0$ are the three lattice vectors in the CCDW phase that form the superlattice with these \textit{in-plane} basis vectors. Here, $a_0$, $b_0$ and $c_0$ parameters follow the lattice vectors as per the hexagonal 1T-Cd$I_2$ type structure. Although phonon softening indicates lattice instability, we do not explicitly compute electron-phonon coupling matrix elements or mode-resolved coupling strengths. Therefore, the present results do not provide a quantitative description of the coupling mechanism and our conclusions are limited to structural and electronic reconstruction.

\textit{Fermi surface nesting (FSN)}: The relationship between Fermi surface nesting (FSN) (refer to Figure \ref{fig-3}) and charge-density-wave formation has been widely discussed in the literature \cite{PhysRevB.107.165115, PhysRevB.77.235104, PhysRevB.105.235104}. In the present work, we analyze the evolution of the electronic structure under lattice distortion. The reconstructed Fermi surface in the CCDW phase arises from strong band folding due to Brillouin zone reduction. We note that our conclusions regarding FSN are based on qualitative analysis of Fermi surface topology. A rigorous assessment would require explicit calculation of the electronic susceptibility $\chi(\mathbf{q})$, which is beyond the scope of the present work. 
The reconstructed electronic structure in 1T-TaS$_2$ exhibits band folding, energy gap opening, and the emergence of localized Ta $5d$ states associated with the CCDW phase. Different structural phases display distinct electronic characteristics arising from the interplay between lattice distortion and electronic reconstruction. Experimental ARPES spectra reflect these reconstructed features, although their detailed intensity distribution can depend on matrix elements and surface conditions. A quantitative modeling of such effects is beyond the scope of the present work.

\begin{figure}
\centering
\includegraphics[width=14.6cm,height=5.6cm]{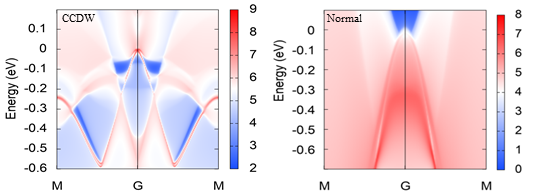}
  \caption{Surface state spectra calculated from wannierTools \cite{MOSTOFI2008685} along high symmetry points \textit{M-$\Gamma$-M} for CCDW and normal phase.}
  \label{fig-5}
\end{figure}

\begin{figure}
\centering
\includegraphics[width=8.6cm,height=7.6cm]{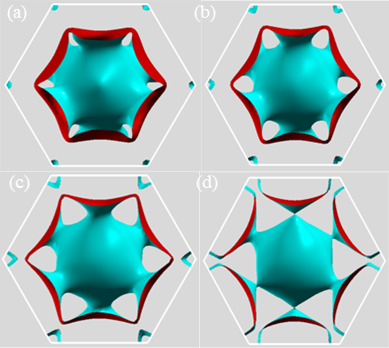}
  \caption{Fermi surfaces at various (a) 0.0, (b) -0.1, (c) -0.2 and (d) -0.3 eV Fermi energy values for the system. The Fermi surface elliptical pockets are showing change in shape with respect to Fermi energy.}
  \label{fig-6}
\end{figure}

\begin{figure*}
\centering
\includegraphics[width=8.5cm,height=9.0cm]{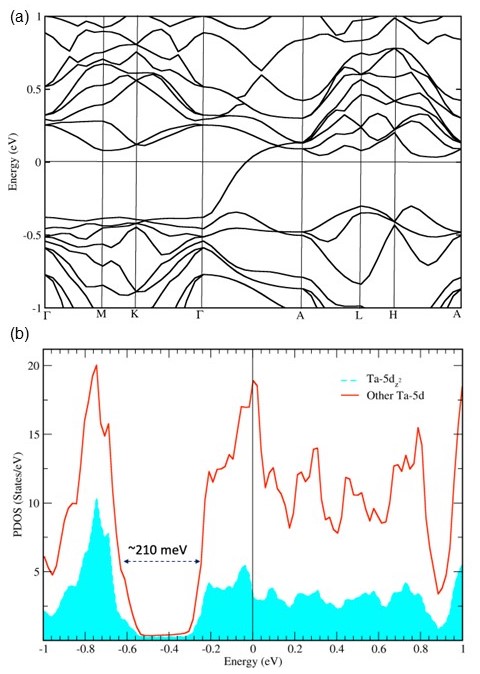}
  \caption{Electronic band structure of the bulk CCDW phase along high symmetry directions (a). Orbital-resolved partial density of states (PDOS) for the CCDW bulk phase, where the blue curve represents the Ta $5d_{z^2}$ orbital and the red curve corresponds to the summed contribution of the remaining four Ta $5d$ orbitals (b).}
  \label{fig-7}
\end{figure*}

\begin{figure*}
\centering
\includegraphics[width=12.5cm,height=8.0cm]{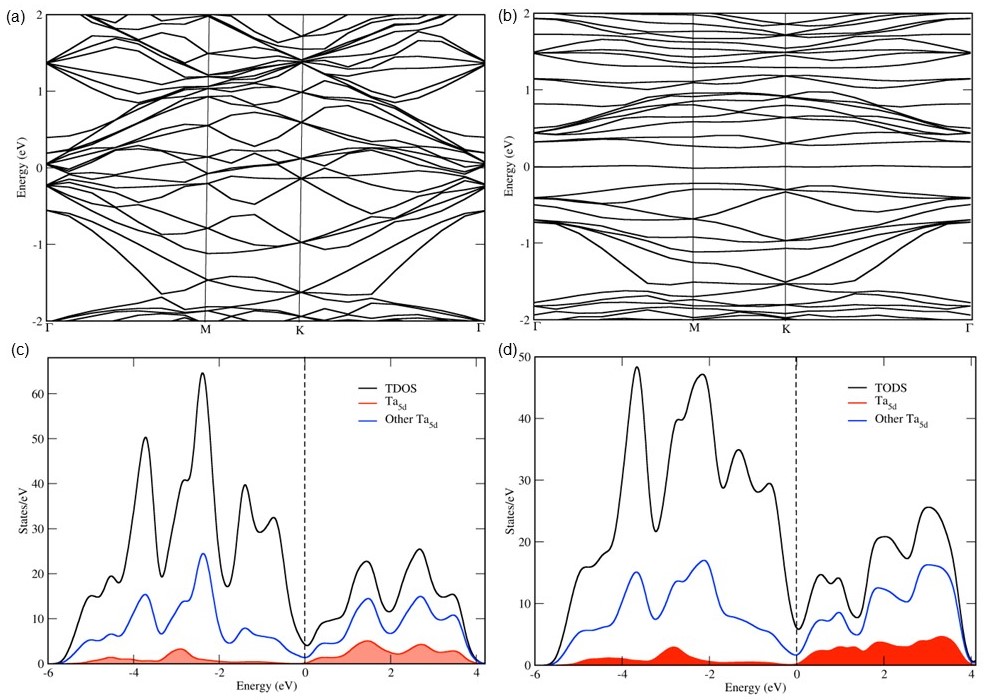}
  \caption{Electronic band structure of the monolayer superlattice undistorted (a) and distorted CCDW phase (b) along high symmetry directions. The band structure is plotted along the standard $\Gamma$–M–K–$\Gamma$ path for a hexagonal Brillouin zone. Orbital-resolved partial density of states for 5d orbitals of 13 Ta atoms in a Star-of-David cluster for undistorted (c) and CCDW phase (d). All band structures and densities of states shown in this figure are obtained from DFT calculations.}
  \label{fig-8}
\end{figure*}

In the present calculations, rev-vdW-DF2 vdW corrections has been included in the structural calculations due to the size and lower symmetry in the distorted superlattice unit cell CCDW phase for better computational accuracy.  We optimize the atomic positions within the superlattice providing slight in-plane displacements randomly within 1-3$\%$ to Ta atoms to break the symmetry. The atoms relaxed to a CCDW phase spontaneously. Thus, the PLD transforms the normal 1T phase into the CCDW phase, leading to SOD cluster formation. For clarity and better understanding, surface states of Wannier bands are plotted in both cases (shown in Figure \ref{fig-5}) and compared with Wannier90 band structure to realize the instability in the CCDW phase. It is obvious to notice the effect of Fermi energy on FSN of the CCDW phases which is shown for various Fermi energies (refer to Figure \ref{fig-6}), where the Fermi surface elliptical pockets are showing change in shape. We have plotted the band structure and partial DOS for bulk (shown in Figure \ref{fig-7}) and monolayer (shown in Figure \ref{fig-8}) phase.

Bulk CCDW states start forming by unit cells that are commensurate with the original 1T phase of Ta$S_2$ bulk structure. The initial cell volume and lattice positions follow high-symmetry structure based undistorted unit cell (P-3m1 space group 164), which is considered as the normal phase at room temperature and no signature of distortions. Small discrepancies between peak positions in the density of states and flat bands in the band structure arise from k-point sampling and broadening used in DOS calculations. 
Figure \ref{fig-7} shows the electronic band structure and partial density of states (PDOS) for the bulk CCDW phase, while the corresponding results for the monolayer are presented in Figure \ref{fig-8}. 

\textit{Periodic lattice distortions in CCDW phase}-~The normal phase 1T-Ta$S_2$ system takes high symmetry state P-3m1 (Space group 164). We performed full wanniersization of Ta d-orbitals using WANNIER90 and check its convergence with DFT band structure for high symmetry phase. Fermi surface has been plotted from the WANNIER90 results and visualized using XCrysDen software \cite{KOKALJ1999176} (shown in Figure \ref{fig-s1}). In the $\sqrt{13}\times\sqrt{13}$ superlattice unit cell, total 13 Tantalum (Ta) atoms and 26 Sulfur (S) atoms present to form the structure. To obtain the commensurate CDW phase from this superlattice undistorted unit cell, we induce periodic lattice distortions into all the Ta atoms centering one Ta atoms 
(shown in Figure \ref{fig-s2}). We checked both bulk ((shown in Figure \ref{fig-s3})) and monolayer ((shown in Figure \ref{fig-8})) cases of CCDW phase and notice the up-gradation in electronic band structures and formation of the \textit{Star of David} (SOD) clusters. The role of Fermi surface nesting (FSN) in charge-density-wave formation has been widely discussed in 1T-TaS$_2$. In the present work, we analyze the evolution of the electronic structure under the commensurate lattice distortion and find that the reconstructed Fermi surface originates primarily from Brillouin zone folding induced by the $\sqrt{13}\times\sqrt{13}$ Star-of-David distortion. We do not explicitly compute the electronic susceptibility $\chi(\mathbf{q})$, and therefore do not claim a quantitative assessment of nesting strength. Instead, our results suggest that apparent nesting-like features in the CCDW phase emerge naturally from band reconstruction rather than acting as an independent driving mechanism.


The CCDW phase can be understood in terms of Brillouin zone reconstruction associated with the $\sqrt{13}\times\sqrt{13}$ superlattice. This reduction of the Brillouin zone leads to band folding and a redistribution of electronic states in momentum space. 

In the undistorted phase, the system exhibits the standard hexagonal Brillouin zone, while the CCDW phase corresponds to a reduced zone rotated by approximately 13.9$^\circ$, consistent with the superlattice periodicity. This reconstruction is illustrated schematically illustrated in the \textit{inset} of Figure \ref{fig-1}. In the normal phase, the Fermi surface exhibits elliptical pockets near the Brillouin zone boundaries. Upon entering the CCDW phase, these features are reconstructed into multiple smaller pockets due to band folding, forming the characteristic $(\sqrt{13}\times\sqrt{13})R13.9^\circ$ pattern.

Moreover, elliptical Fermi surface pockets are marked in normal phase to illustrate the symmetry and evolution of the Fermi surface upon reconstruction. Our first-principles simulation result is consistent where one Ta-\textit{5d} band crosses the Fermi level ($E_F$), forming six oval-like pockets near the Brillouin zone boundaries. Meanwhile, the CDW potential retains the BZ in the CCDW phase and the Fermi surface reconstructs into multiple reduced spots with ($\sqrt{13}$×$\sqrt{13}$)R13.$9^0$ periodicity. Thus, these reconstructed Brillouin zone features reflect the symmetry reduction associated with the CCDW phase and may influence anisotropic electronic responses \cite{Wang2020,acs.nanolett.1c04990,PhysRevB.96.125138}. 

We can notice that Ta atoms are distorted towards the central Ta atoms of the SOD cluster (Figure 2). We find the a-b and a-c interatomic distances that compressed by 4.7$\%$ (4.5$\%$ expt. value) and 3.2$\%$ (2.9$\%$ expt. value), respectively which is in close agreement with experimental report \cite{PhysRevB.56.13757}. Mostly, the Ta atoms distort in-plane direction, so S atoms displace towards out-of-plane direction (Figure 1, bottom right panel) due to the Ta interatomic compression. Figure \ref{fig-s3} shows the band structure of CCDW phase that is different from the bulk phase. Here, an \textit{in-plane} band gap opens at the Fermi level, which is the most noticeable fact. Here, Ta \textit{d-bands} form strongly localized bands up to -1.0 eV the valence band below Fermi level. It is observed that the uppermost p-bands are localized near to -1.48 eV, which is in agreement with experiment \cite{Manzke_1989,PhysRevB.89.155137}. This supports the presence of a lattice-driven electronic reconstruction in the CCDW phase, consistent with a phonon-softening-driven instability.

The reconstructed electronic structure in 1T-TaS$_2$ with the CCDW phase has been extensively studied using both experimental and theoretical techniques. The CDW-induced structural transitions result in the formation of energy gaps, modified band dispersions, and localized states within the gaps. Different CCDW phases exhibit distinct electronic properties and ordering patterns. The interpretation of photoemission data requires accounting for this effect, which can significantly impact the observed spectral features. Furthermore, the electronic structure and CDW phases in 1T-Ta$S_2$ can be manipulated by external perturbations, providing opportunities for potential applications in future electronic devices.


\textit{Conclusions}-~We have investigated the structural and electronic reconstruction associated with the commensurate charge-density wave phase in 1T-TaS$_2$ using density functional theory and Wannier-based modeling. Structural relaxation leads to the formation of $\sqrt{13}\times\sqrt{13}$ Star-of-David distortion, accompanied by significant band folding and Fermi surface reconstruction. Our results establish a direct connection between lattice distortion and the evolution of the electronic structure. In particular, features that resemble Fermi surface nesting arise naturally from band folding rather than serving as an independent driving mechanism. We emphasize that the present study does not include explicit calculations of electronic susceptibility or electron--phonon coupling. Therefore, the results should be interpreted as providing a consistent framework for understanding structural and electronic reconstruction, rather than a complete microscopic theory of CDW formation.

\medskip


\bibliography{main}

\pagebreak
\begin{center}
\textbf{\large Supplementary Materials}
\end{center}
\setcounter{equation}{0}
\setcounter{figure}{0}
\setcounter{table}{0}
\setcounter{page}{1}
\makeatletter
\renewcommand{\theequation}{S\arabic{equation}}
\renewcommand{\thefigure}{S\arabic{figure}}
\renewcommand{\bibnumfmt}[1]{[#1]}
\renewcommand{\citenumfont}[1]{#1}

\begin{figure*} [h]
	\includegraphics[width=5in]{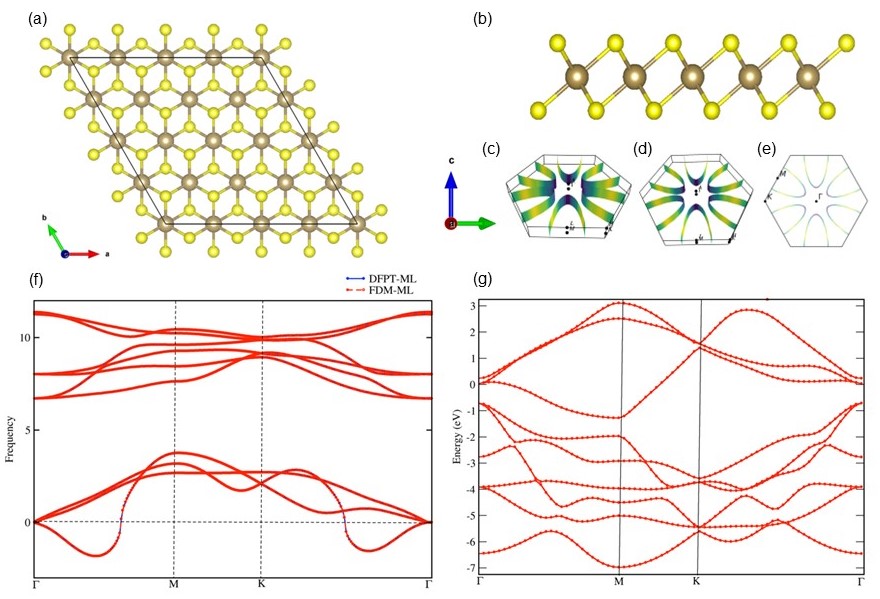}
	\caption{Atomic structures of a 1T-Ta$S_2$ monolayer (a) top view and (b) side view. (c-e) shows the Fermi surface from side view, top view and slice view of the monolayer phase. (f) phonon dispersion plot showing soft phonon modes as negative frequencies. (g) the electronic band structure along high symmetry directions.}
	\label{fig-s1}
\end{figure*}

\begin{figure*} [h]
	\includegraphics[width=7in]{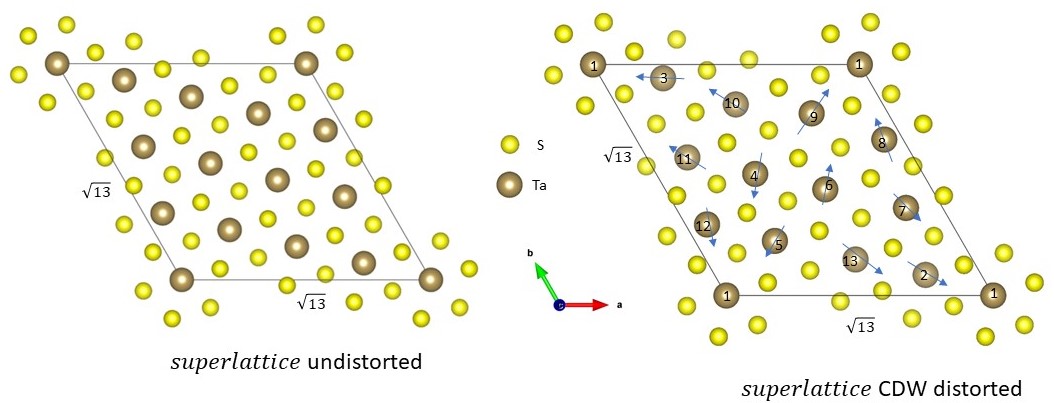}
	\caption{Periodic lattice distortion to obtain CCDW phase from the undistorted superlattice phase.}
	\label{fig-s2}
\end{figure*}

\begin{figure*} [h]
	\includegraphics[width=7in]{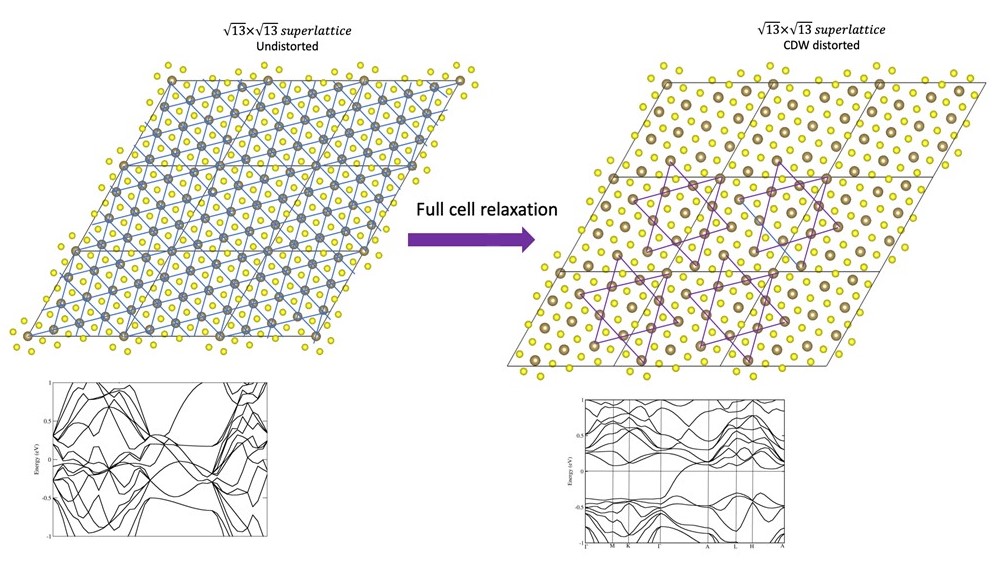}
	\caption{Lattice relaxation for bulk CCDW phase with electronic band structure plots.}
	\label{fig-s3}
\end{figure*}


\end{document}